\documentclass[12pt,preprint,aps,epsfig]{revtex4-1}

\usepackage{graphicx}

\begin{document}
\title{Entropic force in black hole binaries and its Newtonian limits}
\author{Maurice H.P.M. van Putten}
\affiliation{Korea Institute for Advanced Study, Dongdaemun-Gu, Seoul 130-722, Korea}

\begin{abstract}
We give an exact solution for the static force between two black holes at the turning points in their binary motion. The results are derived by Gibbs' principle and the Bekenstein-Hawking entropy applied to the apparent horizon surfaces in time-symmetric initial data. New power laws are derived for the entropy jump in mergers, while Newton's law is shown to derive from a new adiabatic variational principle for the Hilbert action in the presence of apparent horizon surfaces. In this approach, entropy is strictly monotonic such that gravity is attractive for all separations including mergers, and the Bekenstein entropy bound is satisfied also at arbitrarily large separations, where gravity reduces to Newton's law. The latter is generalized to point particles in the Newtonian limit by application of Gibbs' principle to world-lines crossing light cones.
\end{abstract}

\maketitle

\section{Introduction}

Entropy is a central theme in gravity in the presence of black holes, where it is attributed to their event horizon \citep{bek73}. For isolated black holes in thermodynamic equilibrium, the entropy $S_H$ can be calculated to be one fourth of the surface area $A_H$ in Planck units on the basis of its black body radiation properties \citep{haw75b}, 
\begin{eqnarray}
S_H=\frac{1}{4}A_H,
\label{EQN_1}
\end{eqnarray}
where $A_{H}=16\pi M^2$. 

For black hole binaries, the total entropy is a sum of (\ref{EQN_1}) corrected for interactions. In this event, the total entropy can be calculated from the entropy $S_{AH}=\frac{1}{4}A_{AH}$ conform the surface area $A_{AH}$ of the apparent horizon surfaces, defined in terms of the outer most trapped surfaces (e.g.\cite{boo08,nis09}). For a binary of masses $M_i$ $(i=1,2)$ with a separation $a$, we have
\begin{eqnarray}
S_{AH,i}=\gamma_i A_{H,i},
\label{EQN_1b}
\end{eqnarray}
where $\gamma_i=\frac{1}{4} f_i(\xi_1,\xi_2)$, $\xi_i=M_i/a$, includes a form factor $f_i$ that represents a deformation of black hole $i$ by its neighbor $j$ as a function of the separation $a$. The limit (\ref{EQN_1}) of isolated black holes implies $f_i= 1+O(\xi_j)$ in the limit of large $a$. In the approximation of local thermodynamic equilibrium, their temperatures satisfy
\begin{eqnarray}
T_{AH,i} = \left(\frac{\partial S_{AH,i}}{\partial M_i}\right)^{-1} = \left(\gamma_i\frac{dA_{H,i}}{dM_i}+A_{AH,i}\frac{\partial\gamma_i}{\partial M_i}\right)^{-1} \simeq \left(\gamma_i\frac{dA_{H,i}}{dM_i}\right)^{-1}
\label{EQN_2}
\end{eqnarray}
in the limit of large $a$. The binding energy $U$ between the two black holes can now be calculated using Gibbs' principle (e.g. \cite{neu09}) from the entropy difference $dS_{AH,i} =A_{H,i} d\gamma_i$ between two static configurations at separations $a$ and $a+da$ with otherwise the same total mass energy. By (\ref{EQN_2}), we have $T_{AH_i}dS_{AH_i}\simeq A_{H,i}\left(\frac{dA_{H,i}}{dM_i}\right)^{-1}d\ln \gamma_i$,  and hence with (\ref{EQN_1}), $-dU = \left(T_{AH} dS_{AH}\right)_1+ \left(T_{AH} dS_{AH}\right)_2$ integrates to
\begin{eqnarray}
U\simeq - \frac{M_1}{2} \ln \gamma_1 - \frac{M_2}{2} \ln \gamma_2.
\label{EQN_4}
\end{eqnarray}

In this paper, we shall derive exact expressions for (\ref{EQN_1b}) for black hole binaries at their turning points. The results define the static force between two black holes at large and small separations from (\ref{EQN_4}), including mergers to a single black hole. Our fully nonlinear solution covering the full range of separations, from large down to the merger phase including the associated entropy jumps, goes beyond a recent dimensional analysis of \cite{mor11}. The equations of motion, that includes the inertia of the black holes, will be derived by extending the Hilbert action with additional boundary terms to account for the presence of apparent horizon surfaces.

Our exactly solvable example demonstrates some general {\em entropic constraints} that any entropic theory of gravitation must satisfy: monotonicity of entropy, i.e., a generally attractive force of gravity and the Bekenstein entropy bound \cite{bek81} at all separations. 

In \S2, Newton's law is shown to appear in the leading order expansion of (\ref{EQN_4}), using an exact solution for the time-symmetric data of two black holes by solving the Hamiltonian energy constraint. In \S3, the fully nonlinear solution to (\ref{EQN_4}) is given, using both perturbative and numerical solutions. In \S4, the Hilbert action is extended to include apparent horizon surfaces that, by an adiabatic variational principle, is shown to recover Newton's law at large separations. The results show entropic constraints that any entropic theory must satisfy, that point to a generalization to Newton's law for point particles, discussed in \S5 by application of Gibbs' principle to light cones with a reflection on the Verlinde approach on holography using time-like screens\citep{ver10}. Some conclusions are given in \S6.

\section{Newton's law from apparent horizon surfaces}

The time symmetric data for two black holes of mass $M_1=M$ and $M_2=m$ at coordinate positions $p$ and, respectively, $q$ \citep{bri63,lin63,mis63,bri63,coo01} are described by the conformal factor
\begin{eqnarray}
 \phi=1+\psi_M(p)+\psi_m(q),
 \label{EQN_ID}
 \end{eqnarray}
in a conformally flat metric 
\begin{eqnarray}
h_{ij}=\phi^4\delta_{ij}
\end{eqnarray}
as an exact solution to the Hamiltonian energy constraint, where $\psi_M(p)=\frac{M}{2|r-p|}$ and $\psi_m(q)=\frac{m}{2|r-q|}$ in terms of the Green's function $\frac{1}{|p-r|}$ of the Laplacian of the flat metric $\delta_{ij}$. To leading order in the separation $a=\left|p-q\right|$, the deformation of black hole $i$ due to its neighbor $j$ is a function of $\xi_j=M_j/a$ only $(i\ne j)$, i.e.,
\begin{eqnarray}
A_{AH,i} \simeq 16\pi M_i^2f\left(\xi_j\right).
\label{EQN_5}
\end{eqnarray}
To calculate $f$, consider a spherical coordinate system $(\rho,x,\varphi)$, $x=\cos\theta$ for the flat three-metric $\delta_{ij}$ with origin at the location $q$ of $M_2=m$. The surface area $A=A_{AH,2}$ of the apparent horizon surface,
\begin{eqnarray}
A(\rho) = 2 \pi \int\phi^4 \rho^2 dx,
\label{EQN_N0}
\end{eqnarray}
corresponds to leading order to the extremum of the area of the coordinate surfaces of constant $\rho$,
\begin{eqnarray}
A^\prime(\rho)=0,
\end{eqnarray}
where
\begin{eqnarray}
A(\rho) = 2 \pi \int\phi^4 \rho^2 dx \simeq 4\pi\left[1+\frac{m}{\rho}+\frac{M}{a}+\frac{m^2}{4\rho^2} + \frac{mM}{2\rho a} +\frac{M^2}{4a^2}
\right]\rho^2.
\label{EQN_N1}
\end{eqnarray}
The coordinate radius of the perturbed event horizon satisfies $\rho \simeq \frac{m}{a}\left(1-\frac{M}{2a}\right)$, whereby the surface area $A_{H,2}=16\pi m^2$ of the event horizon of an isolated Schwarzschild black hole of mass $m$ changes to
\begin{eqnarray}
A_{AH,2} = 16\pi m^2f(\xi),~~f(\xi)=1+\frac{M}{a} +\cdots
\label{EQN_A2}
\end{eqnarray}
with dots referring to higher order terms in the perturbative expansion. 

Newton's law is immediately apparent from (\ref{EQN_4}) and (\ref{EQN_A2}) in representing the leading order perturbation of the areas of the apparent horizon surfaces owing to black hole-black hole interactions. Given our fully nonlinear model, we next turn to a solution for all separations, including the merger phase of two black holes. 

\section{Entropy creation in a merger}

Let $T$ denote a trapped surface: a two-dimensional closed surface in a Cauchy surface $\Sigma$, whose outgoing (and ingoing) future directed null-geodesics orthogonal to $T$ have negative (positive) rates of expansion \citep{poi07}. These outgoing null normals $k^b$ can be expressed as the sum of the unit (time-like) $n^a$ normal to $\Sigma$ and the unit (spacelike) $s^i$ normal to $T$ in $\Sigma$ \citep{wal91}
\begin{eqnarray}
k^b = n^b + s^b.
\label{EQN_k}
\end{eqnarray}
The causal structure of $T$ is such, that at each point of $T$, $k^b$ appears to be directed along the past light cone of an observer passing through $T$. In a 3+1 representation of the metric with three-metric $h_{ij}$ on $\Sigma$ and extrinsic curvature $K_{ij}$, the {\em apparent horizon surface} is defined as the outermost marginally trapped surface - a ``frozen light cone," that neither expands nor contracts. The equation for a marginally trapped surface is $q^{ab}\nabla_ak_b=0$ \citep{wal91}, where $q_{ab}=g_{ab}+n_an_b-s_as_b$ denotes the metric tensor induced in $T$, i.e.,  \cite{bre88,yor89,wal91,coo92,coo01,tho07}
\begin{eqnarray}
\Theta \equiv \nabla^is_i+K_{ij}s_is_j-K^2=0.
\label{EQN_TH01}
\end{eqnarray}
For time-symmetric data, the extrinsic curvature tensor vanishes, whereby (\ref{EQN_TH01}) reduces to \index{time-symmetric initial data}
\begin{eqnarray}
\Theta \equiv \nabla^is_i=0.
\label{EQN_TH1}
\end{eqnarray}

To solve (\ref{EQN_TH1}), we use spherical coordinates $(\rho,x,\varphi)$, $x=\cos\theta$, (\ref{EQN_TH1}) and obtain a nonlinear problem in
\begin{eqnarray}
s_i=(\phi^2\cos\lambda,\frac{\rho\phi^2}{\sqrt{1-x^2}}\sin\lambda,0), \tan\lambda(x)=-\sqrt{1-x^2}f^\prime(x), \rho(x)=\rho_0e^{f(x)},
\label{EQN_TH2}
\end{eqnarray}
where azimuthal symmetry $(\partial_\phi=0$) is used for two black holes along the $z-$axis with coordinate positions $p$ and $q$. In this axisymmetric configuration, the horizon is a surface of revolution with area 
\begin{eqnarray}
A=2\pi \int_{-1}^1 \phi^4\rho^2\frac{dx}{\cos\lambda},
\end{eqnarray}
where integration is over one or two event surfaces depending on the separation between the two black holes. The $\lambda$ function satisfies the ordinary differential equation given by
\begin{eqnarray}
\lambda^\prime + 4\rho\frac{\partial_\rho\phi}{\phi}+2 + 4\tan\lambda \frac{\partial_\theta \phi}{\phi} + \frac{\tan\lambda}{\tan\theta}=0
\end{eqnarray}
with $\rho^\prime = \rho\tan\lambda,$ $A_H^\prime =2\pi\frac{\phi^4\rho^2}{\cos\lambda}\sin\theta$ subject to $\lambda=0~~(\theta=0,\pi),$ from which both analytic and numerical solutions can be derived.

\begin{figure}[h]
\centerline{\includegraphics[width=120mm,height=80mm]{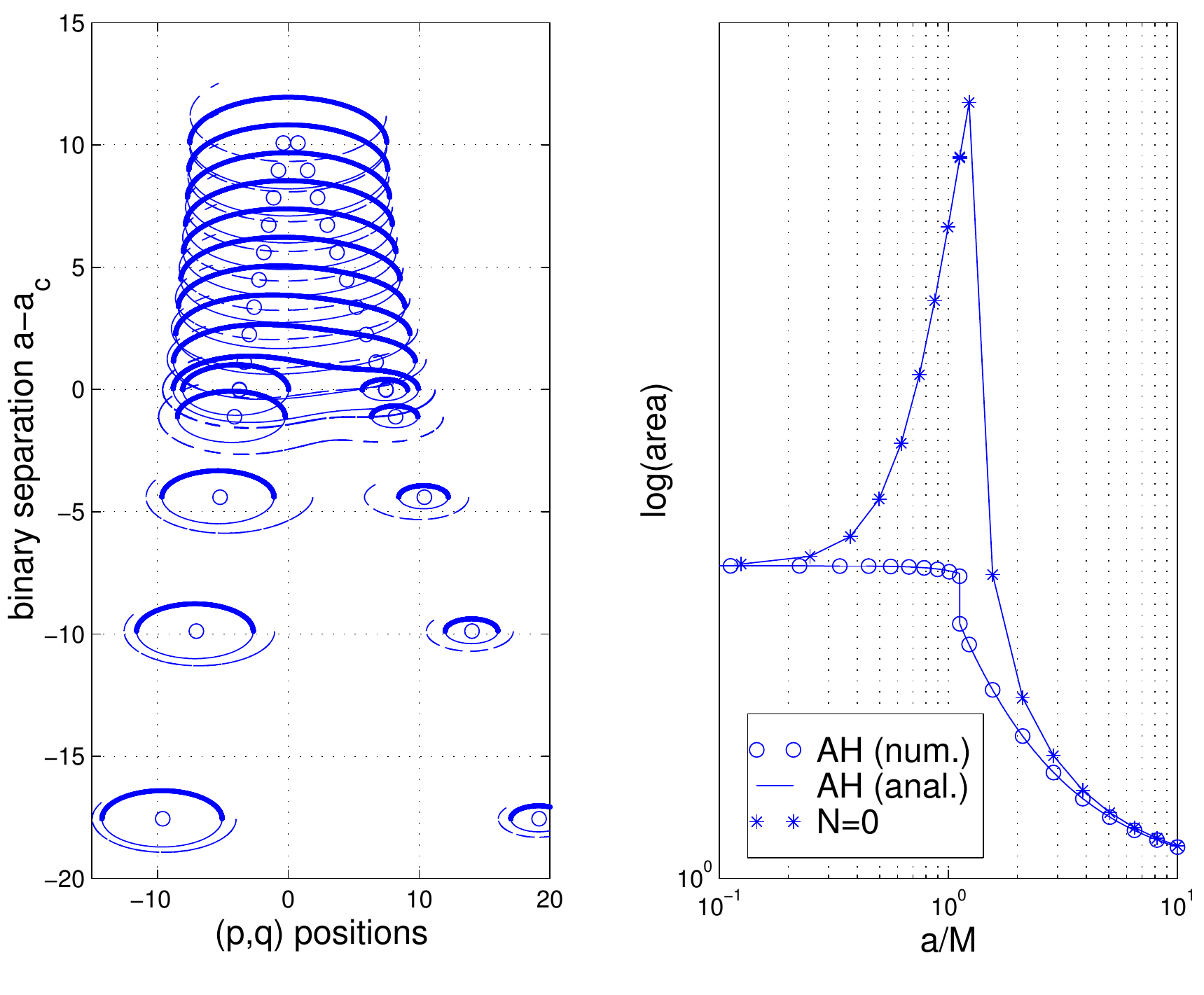}}
\centerline{\includegraphics[width=120mm,height=60mm]{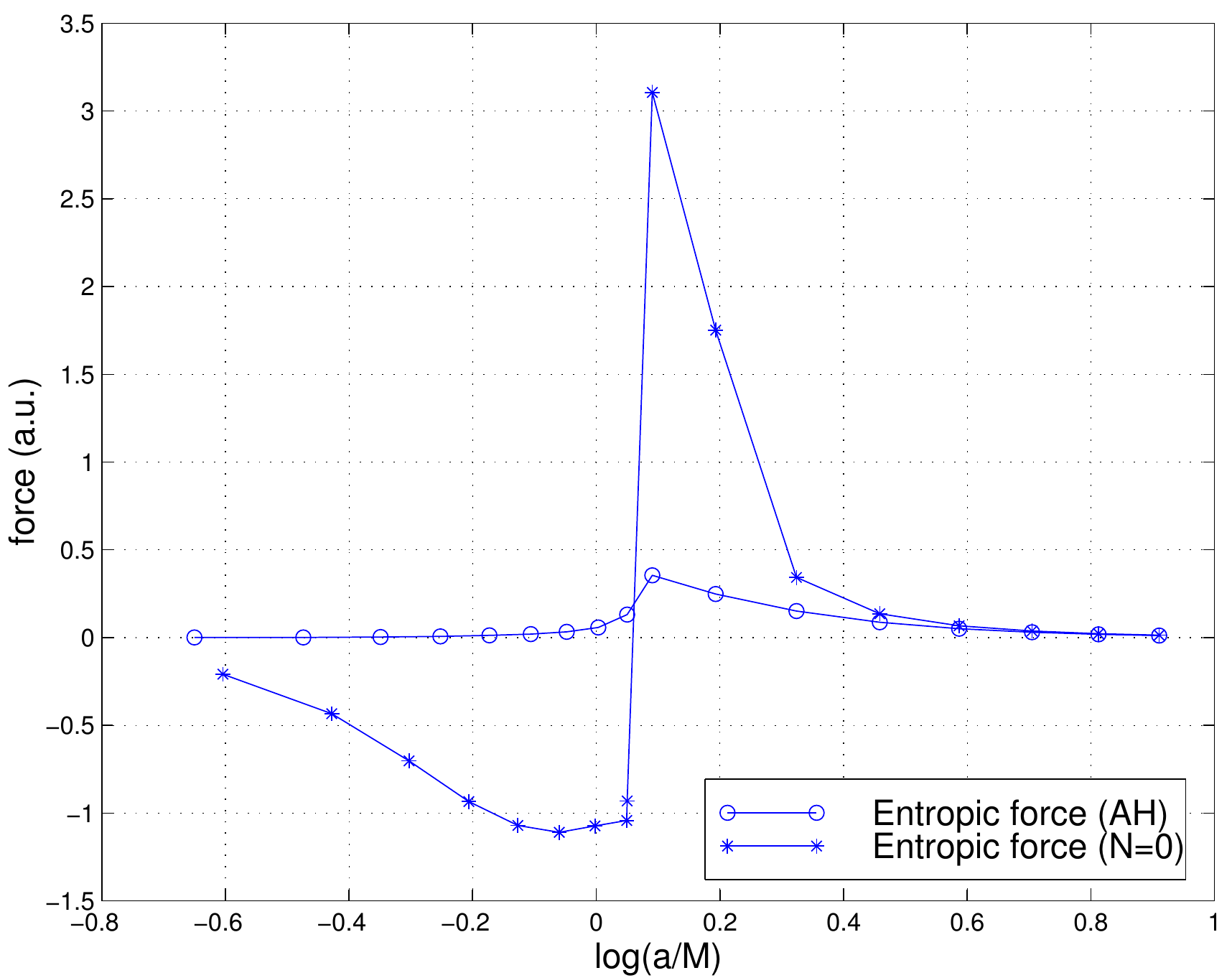}}
\caption{{\small ($Top~left.$) Snapshots of the apparent horizon surface ({\em thick arcs}) and surfaces of infinite redshift ({\em thin arcs}, $N=0$) in a black hole binary as a function of separation in time-symmetric data for a mass ratio 2:1. ($Top~right.$) The surface area of the apparent horizon surface(s) is shown normalized to $A_H=16\pi (M^2+m^2)$ in the limit as the separation approaches infinity, computed numerically ({\em circles}) and by high order inner Taylor expansion and an outer Laurent expansions ({\em continuous line}) in the separation distance. For a mass ratio of 2:1, we have $A_H$(normalized)$\le1.8$. A common apparent event horizon forms at the bifurcation point $a/M=14.9$ when $m/M=0.5$ ($a/M=15.3$ when $m/M=1$; \cite{cad73}). The surface area of surfaces of constant redshift, $A_{N=0}$, changes direction in slope across the bifurcation point. ($Bottom$) The entropic force following (\ref{EQN_4}) satisfies monotonicity in the area as a function of separation, giving an attractive entropic force throughout, whereas the surfaces $A_{N=0}$ with the same entropy density do not.}}
\label{FIG_QC1}
\end{figure}

Fig. \ref{FIG_QC1} shows the apparent horizon surfaces along with surfaces of infinite redshift in a black hole binary for a sequence of separation distances. The entropic force is calculated following surface integrals of $T_{AH}\frac{\partial^2S_{AH}}{\partial a \partial A_{AH}}$, to include the non-uniform temperatures over the apparent horizon surfaces. Because the entropy of the apparent horizon surfaces satisfies strict monotonicity for all separations, it produces an entropic force which is always attractive. This result may be compared with the entropic force that one would infer from attributing a constant entropy per unit surface area of surfaces of constant redshift. Surfaces $A_{N=0}$ of vanishing redshift does not satisfy monotonicity across the bifurcation. Following the formation of a common event horizon surface in the merger of the two black holes, the entropic force becomes repulsive. This observation serves to illustrate that holography using constant redshift surfaces will have a variable Bekenstein-Hawking entropy density, that will be non-maximal and less than 1/4.

The outer and inner expansions of the surfaces of the apparent horizon surfaces derive from the Laurent and Taylor series in the dimensionless separation $p/m$ for $u=m/p<1$ and $v=p/m<1$, following the Legendre expansions for $\phi$:
\begin{eqnarray}
\Phi_o = 1+\frac{M}{2\rho}+\frac{m}{2p}\Sigma\left(\frac{\rho}{p}\right)^lP_l(x),~~\Phi_i = 1+\frac{1}{2}\Sigma\frac{I_l}{\rho^{l+1}}P_l(x),
\label{EQN_THoi}
\end{eqnarray}
where, in the center of mass frame, we define the moments 
\begin{eqnarray}
I_n=mq^n\left[1+(-1)^n\left(\frac{m}{M}\right)^{n-1}\right].
\end{eqnarray}
Perturbative solutions to (\ref{EQN_TH1}) now follow from the {\em Ans\"atze} $\rho_o=\frac{M}{2}e^{f_o(x)},$ and $\rho_i=\frac{M+m}{2}e^{f_i(x)}$ with each $f_{o,i}(x)$ expanded in Legendre polynomials. We can thus expand $f_o$ to 6th order and $f_i$ to 10th order using symbolic computation, leading to (\ref{EQN_Ao}) and (\ref{EQN_Ai}).

The series expansion of the area of both apparent horizons is\index{apparent horizon surface!area}
\begin{eqnarray}
A_o=16\pi (M^2+m^2)+16\pi Mm\frac{M+m}{a}+4\pi M^2\frac{m^2}{a^2}\left(2-\frac{M^8}{280m^4a^4}-\frac{m^8}{280M^4a^4}\right)
\label{EQN_Ao}
\end{eqnarray}
when $a>a_c$ and of the common horizon envelope is
\begin{eqnarray}
A_i=16\pi (M+m)^2\left[1-\frac{32}{35}\hat{I}_2^2\right].
\label{EQN_Ai}
\end{eqnarray}
when $a<a_c$, where $\hat{I}_2=I_2/(M+m)^3$. We note that (\ref{EQN_Ao}) are asymptotic expansions in the mass ratio $m/M$. Here $a_c$ as a function of $m/M$ is a critical separation distance \cite{bri63,cad73}, representing the emergence of a common horizon envelope\index{common horizon envelope} in the slow merger of two black holes. Numerically, the jump $\Delta A_H$ in the total horizon surface area across the bifurcation point scales approximately linearly with the normalized moment of inertia $\frac{I_2}{(M+m)^3}$ and, as a function of  the bifurcation parameter $x=\frac{\delta A_H}{A_H}=1-\frac{M^2+m^2}{(M+m)^2}$, is found to closely satisfy the power laws
\begin{eqnarray}
\frac{\Delta A_H}{A_0} \simeq 0.2560 x^{\frac{3}{2}},~~\frac{q_c}{M}\simeq \frac{1}{2}+0.4224x^\frac{2}{3}
\end{eqnarray}
in the range of $0<x<0.5$, where $q_c$ denotes the critical distance of the black hole of mass $m$ to the center of mass of the binary.

When the common horizon envelope forms, (\ref{EQN_Ai}) defines an energy $\delta E=T_H\delta S_H\simeq \frac{16}{35}(M+m)\hat{I}_2^2$ that is distributed in tidal dissipation in the event horizon and in emissions in gravitational radiation. Since $a< (M+m)/2$ in this phase, $\delta E < \frac{\mu^2}{35(M+m)}$ represents a bound on the maximal energy that can be radiated. For $M=m$, we have in particular\index{gravitational radiation}
\begin{eqnarray}
\delta E < \frac{M}{280} = 0.36\% M,
\label{EQN_GW}
\end{eqnarray}
consistent with numerical estimates of about $0.2\%$ M \cite{ann93}. The smallness of the gravitational wave output (\ref{EQN_GW}) is illustrative for the regularization by general relativity of the singular behavior in the Newtonian attraction between two point particles by the formation of event horizons. 

\section{Newton's law from an adiabatic variational principle}

To derive equations of motion from an entropic principle, we must go beyond Gibbs' principle and include inertia. To begin, we consider the Hilbert action
\begin{eqnarray}
S=\frac{1}{16\pi}\int R d^4x 
\label{EQN_H0}
\end{eqnarray}
in terms of the Ricci three-scalar $R$. Here, we may add a boundary term $\frac{1}{8\pi}\int_\partial K$ in case space is not asymptotically flat \citep{haw96}. In 3+1, the line element becomes 
\begin{eqnarray}
ds^2=-N^2dt^2+h_{ij}(dx^i+\beta^i)(dx^j+\beta^j),
\end{eqnarray} 
where $h_{ij}$ denotes the three metric of a foliation given by a lapse function $N$ and shift vectors $\beta^i$ with extrinsic curvature tensor $\dot{h}_{ij} = D_i\beta_j+D_j\beta_i-2NK_{ij}$, using the covariant derivative $D_i$ and the scalar Ricci tensor $^{(3)}R$ induced by $h_{ij}$.  In 3+1, we have $S=\frac{1}{16\pi}\int \left( ^{(3)}R+K:K-K^2\right) \sqrt{h}Nd^3 xdt$. Variation with respect to the non-dynamical variable $N$ gives rise to the Hamiltonian energy constraint $R-K:K+K^2=0.$ Without loss of generality, we may further consider a conformal factorization 
\begin{eqnarray}
h_{ij}=\phi^4 g_{ij}
\end{eqnarray}
with conformal factor $\phi$ normalized by $\sqrt{g}=1$, whereby $\sqrt{h}=\phi^6$. We thus have the density $R\sqrt{h}=\left[\phi^{2}R(g)-{8}{\phi}\Delta \phi\right]\sqrt{g}$, whereby (\ref{EQN_H0}) becomes
\begin{eqnarray}
S=\frac{1}{16\pi}\int\left(\phi^{2}R(g)-8\phi D^iD_i\phi +\phi^6 (K:K-K^2)\right) N d^3x dt,
\label{EQN_H1}
\end{eqnarray}
where $D_i$ is the covariant derivative of $g_{ij}$. It has been recognized that (\ref{EQN_H1}) is open to potentially additional boundary terms to generate an action which is quadratic in first derivatives \citep{lan75} and in obtaining finite Euclidian actions for black hole spacetimes \citep{gib77}. 

By (\ref{EQN_ID}), the Lagrangian in (\ref{EQN_H1}) hereby reduces to ${\cal L}=\int_{\theta\ge0} (-\phi\Delta\phi+3\phi^4\dot{\phi}^2)d^3x$ in the gauge with lapse $N=1$ with integration over the singularity-free region outside the apparent event horizons, $\theta\ge0$, where $\theta$ denotes the divergence of null geodesics \citep{yor89}. We observe that ${\cal L}$ now contains no interaction in response to time symmetric data ($\Delta\phi\equiv 0$ and $\dot{\phi}\equiv0$), whereby the black holes will not start to move. This contradicts the entropic force (\ref{EQN_4}) as it arises from the same initial data. 
 
We next introduce boundary terms in (\ref{EQN_H1}) to incorporate entropy in spacetimes with well-separated black holes. In doing so, we encounter the fact that the Einstein equations, as they arise from (\ref{EQN_H0}) and (\ref{EQN_H1}), are mixed elliptic and hyperbolic: elliptic in regards to gravitational attraction and entropy associated with apparent horizon surfaces and hyperbolic in regards to gravitational wave motion. The conformal factorization (\ref{EQN_H1}) allows us to introduce boundary terms for obtaining quadratic expressions in the first derivates of the conformal scale factor independently of the hyperbolic degrees of freedom associated with wave motion. 

We define the inertial boundary term
\begin{eqnarray}
{\cal M}=\frac{1}{2\pi}\int_{\theta=0} \left[ N \phi D_n\phi - \phi^2 D_nN \right]
\label{EQN_M}
\end{eqnarray}
associated with the apparent horizon surfaces $\theta=0$ as it appears in Green's identity ${\cal M}=\int \left[ N \phi D^iD_i\phi  - \phi^2\Delta_gN  \right] d^3x + \int  N D^i\phi D_i\phi  dx^3$. Subtracting ${\cal M}$ from the Lagrangian in (\ref{EQN_H1}) gives
\begin{eqnarray}
S=\int {\cal L} Ndx^2 dt
\end{eqnarray}
with
\begin{eqnarray}
{\cal L}=\phi^{2}R(g)-16 \phi D^i D_i\phi - 8D^i\phi D_i\phi +8\phi^2N^{-1}\Delta_gN +\phi^6 (K:K-K^2).
\label{EQN_L}
\end{eqnarray}

The boundary term (\ref{EQN_M}) can be understood by considering an isolated black hole, described by $\phi=1+\frac{M}{2r}$ in the gauge $N=\frac{2-\phi}{\phi}$, whereby $\theta=0$ and $N=0$ coincide. In this event, we have
\begin{eqnarray}
{\cal M} = M
\label{EQN_M1}
\end{eqnarray}
The event horizon of an isolated Schwarzschild black hole corresponds to the minimum area $A(r)=\int \phi^4d\Sigma = 4\pi r^2\phi^2(r)$ at the coordinate radius $R_g=\frac{M}{2}$, where $\phi=2$ and $A(R_g)=16\pi M^2$. It represents a turning point in view of a M\"obius symmetry in the radial coordinate representing the two sheet embedding of the (exterior) Schwarzschild spacetime. In the same gauge, we find a horizon surface gravity $g_H=dN/ds=-\phi^{-2}D_nN$ on $N=0$. Since $T_H=\frac{1}{2\pi}g_H$ denotes the Hawking-Unruh temperature of the event horizon, the addition to the action in (\ref{EQN_H1}) satisfies
\begin{eqnarray}
{\cal M} = \frac{1}{2}\int T_H \phi^4 d\Sigma= \frac{1}{2}T_H A_H = 2S_HT_H,
\label{EQN_M2}
\end{eqnarray}
where $S_H=\frac{1}{4}A_H$ denotes the Bekenstein-Hawking entropy. Thus, the boundary term (\ref{EQN_M}) in (\ref{EQN_H1}) represents inertia (\ref{EQN_M1}), setting a correlation (\ref{EQN_M2}) between temperature and entropy. 

Gravitational attraction is defined by the elliptic part of Lagrangian. With trapped surfaces, it is subject to entropic considerations, here described by the conformal scale factor $\phi$. It can be seen by analyzing the interaction between two black holes in the Newtonian limit of absolute time, described by the uniform lapse function $N\equiv1$. In the conformally flat, quadratic approximation to (\ref{EQN_L}) in $\phi$, this approximation gives
\begin{eqnarray}
-S_Q=\frac{1}{2\pi}\int \int_{\theta\ge0} \left[ (\partial\phi)^2+ 3\dot{\phi}^2\right] d^3 x dt.
\label{EQN_LQ2}
\end{eqnarray}
The first integral is represented by a Lagrangian comprising the three surface integrals,
\begin{eqnarray}
\frac{1}{2\pi} \int_{\theta=0} \psi\partial_n\psi= \frac{1}{2\pi}\int_{\theta=0}\left[ \psi(p)\partial_n\psi(p)+ \psi(q)\partial_n\psi(q) + 2\psi(p)\partial_n\psi(q)\right]d\Sigma.
\end{eqnarray}
Here, and in what follows, we use the short-hand $\psi(p)=\psi_M(p)$ and $\psi(q)=\psi_m(q)$ (cf. \ref{EQN_ID}).
In the adiabatic limit, these integrals are evaluated at constant surface area for each component of the apparent horizon surfaces. In the approximation of spherical symmetry, their integrands effectively depend only on the radius of $\theta=0$, whereby preserving $A_H$ implies that the effective masses $M^\prime=\frac{1}{2\pi}\int_{\theta=0}\psi(p)\partial_n\psi(n)d\Sigma$ and $m^\prime=\frac{1}{2\pi}\int_{\theta=0}\psi(q)\partial_n\psi(q)d\Sigma$ (each over one component of $\theta=0$) are constant and do not partake in the variational principle. The mixed term satisfies
\begin{eqnarray}
 \int_{\theta\ge0}\partial^i\psi(p)\partial_i\psi(q) =\left(\int_{|r-p|=\frac{M}{2}}+\int_{|r-q|=\frac{m}{2}}+\lim_{R\rightarrow\infty}\int_{|r|=R}\right)\psi(p)\partial_n\psi(q)d\Sigma.
\end{eqnarray}
Since $\psi(p)=1$ on $|r-p|=\frac{M}{2}$ and $\psi(q)=1$ on $|r-q|=\rho=\frac{m}{2}$, we have
$\int_{|r-p|=\frac{M}{2}}\psi(p)\partial_n\psi(q)=\int_{|r-p|=\frac{M}{2}}\partial_n\psi(q) = \int_{|r-p|\le\frac{M}{2}} \Delta\psi(q)=0$, and so
\begin{eqnarray}
\frac{1}{\pi} \int_{|r-q|=\rho} \psi(p)\partial_n\psi(q)\simeq\frac{M^\prime m^\prime}{|p-q|},
\end{eqnarray}
while $\lim_{R\rightarrow\infty}\int_{|r|=R}\psi(p)\partial_n\psi(q)=0$. The inertial term in the Lagrangian (\ref{EQN_LQ2}) is
\begin{eqnarray}
\frac{3}{2\pi}\dot{p}^i\dot{p}^j\int_{\theta\ge0}\partial_{p^i}\psi\partial_{p^j}\psi d^3 x 
=\frac{1}{2\pi}\dot{p}^2\int_{\theta\ge0} (\partial\psi)^2d^3 x = \frac{M^\prime}{2}\dot{p}^2,
\label{EQN_SN2}
\end{eqnarray}
and similarly for the second mass integral associated with $q^i$. We also encounter a velocity cross-correlation 
\begin{eqnarray}
\dot{p}^i\dot{q}^i\frac{1}{\pi}\int_{\theta\ge0} \partial_{p^i} \psi \partial_{q^j} \psi = \dot{p}^i\dot{q}^j\frac{1}{\pi}\int \partial_i \psi(p)\partial_j \psi(q) 
= O\left(\frac{|\dot{p^i}||\dot{q^i}|M^\prime m^\prime}{|p-q|}\right).
\end{eqnarray}
It follows that the adiabatic limit of the quadratic approximation to (\ref{EQN_L}) obtains the action for a binary of point particles in classical mechanics,
\begin{eqnarray}
- S_Q = \int \left[ \frac{1}{2}M\dot{p}^2 + \frac{1}{2}m\dot{q}^2 + \frac{Mm}{|p-q|}\right]dt,
\label{EQN_NT}
\end{eqnarray}
where we dropped the primes over the masses. The variational principle by way of the $p,q$ trajectories is such that the associated scalar field is perturbed globally, i.e., variations $\delta p$, $\delta q$ carry along global variations $\phi_p\delta p$ and $\phi_q\delta q$. This Newtonian ``frozen field" approach (apart from global translations) is evidently distinct from local variations in $\phi$ as used in the deriving equations of motions for $\phi$ as a field variable, that would otherwise give rise to an elliptic equation $3\ddot{\phi}+\Delta \phi=0$.

The derivation leading to the Newtonian limit (\ref{EQN_NT}) shows that the gravitational attraction between two black holes is defined by a Lagrangian obtained by integration of $\phi$ over $\theta\ge0$, which preserves a finite distance away from the singularities at $p^i(t)$ and $q^i(t)$. This regularization by apparent horizon surfaces is a practical manifestation of cosmic censorship, by which black hole interactions are extremely smooth in contrast to the singular behavior in the Newtonian interaction between point particles. The regularized evolution in the slow motion (elliptic) ``frozen" field approximation (\ref{EQN_ID}) defines a functional $S=S(p,q)$ (with the fully nonlinear $\phi^4\dot{\phi}^2$ kinetic energy term or in the quadratic approximation (\ref{EQN_LQ2}) on the basis of tracking $\theta=0$ as a function of $(p,q)$ by numerical evaluation.

Our approach results in a coupled system of equations for gravitational waves, encoded in $g_{ij}$, for a completely regularized  gravitational attraction. Thus, entropic considerations are representative for the elliptic part of the otherwise mixed elliptic-hyperbolic structure of general relativity, giving a separation of wave motion and inertial motion closely related to the separation of wave motion and evolution of causal structure as in the Riemann-Cartan formulation of general relativity \citep{pir56,pir57,van96}. 

\section{Newton's law for point particles} 

Our exact solution of the entropic force for a black hole binary satisfies {\em monotonicity of entropy}, giving rise to an attractive gravity force for all separations. Since the entropy of the apparent horizon surfaces approaches the sum of the entropy (\ref{EQN_1}) of the two black holes in the limit of large separations, the {\em Bekenstein entropy bound} $S\le 2\pi E R=\frac{E}{2R}A$ \citep{bek81}, where $E$ represents the total energy enclosed within a sphere of area $A=4\pi R^2$, is automatically satisfied.

It is instructive to consider these two entropy constraints in the alternative approach, based on scaling arguments, to Newton' law for point particles that are not black holes as an entropic force using time-like holographic screens \cite{ver10} following \cite{tho93,sus94}. Already we noted that in the merger phase, this proposal gives rise to a repulsive gravitational force if endowed with an entropy density of 1/4. This result shows that the entropy surface density on time like holographic screens will be different for a generally attractive force to arise. A model for calculating the two-dimensional entropy distribution on the screen, in its dependence on the mass distribution within and outside, is an open problem, however, that will be relatively small for large screens in view of the above mentioned Bekenstein entropy bound.

Since null-surfaces are surfaces of causal separation, whether they are event horizons of black holes or extended light cones, it is reasonable to postulate that $S=\frac{1}{4}A$ continues to hold. To be precise, $S$ is hereby equal to the area of the disk enclosed by surfaces of constant phase, i.e., of their projection of  onto the equatorial plane. It gives rise to a correlation
 \begin{eqnarray}
TS=\frac{1}{2}M
\label{EQN_ST}
\end{eqnarray}
similar to (\ref{EQN_1}-\ref{EQN_2}) (with an additional factor $\cos\lambda$, $\sin\lambda=a/M$ on the right hand side for rotating black holes). Upon  restoring dimensionful units, the Bekenstein entropy bound hereby becomes
\begin{eqnarray}
k_BT \tau  \ge \frac{1}{2}\hbar
\label{EQN_UN}
\end{eqnarray}
as an uncertainty relation for the thermal energy $k_BT$ on a two dimensional screen, where $\tau=2\pi t$ is the uncertainty in time associated with its circumference $2\pi R$. 

For particles that are not black holes, therefore, it appears natural to consider their entropic interactions in terms of entropy associated with their future directed light cones, generated by null-geodesics emanating from their world-lines. Their change in entropy now becomes equivalent to a change in area due to gravitational lensing. In the case at hand, there are two relevant surface areas that we may consider: $A_0$ of the equatorial disk contained within a sphere of surface area $4\pi r^2$ and $A_1$ of a wave-front of constant phase. In a Newtonian approximation to the Schwarzschild metric,  we have $\alpha^{-2}=\left(1-\frac{2m}{r}\right)^{-1} \simeq 1+\frac{m}{r}$ and so
\begin{eqnarray}
A_0(r,m)=2\pi \int_{0^+}^r r^\prime ds = 2\pi \int_{0^+}^r r^\prime\left(1+\frac{m}{r^\prime}\right) dr^\prime = \pi r^2 + 2\pi m r,
\label{EQN_A1}
\end{eqnarray}
where $0^+$ refers to neglecting the Schwarzschild radius of the particle. 
For a given $r$ that, in the Schwarzschild line-element fixes the surface area of a sphere centered at the origin, (\ref{EQN_A1}) demonstrates a linear increase of the projected surface area $A_0$ with $m$.  

Following \citep{jac95}, we may use lensing to derive (\ref{EQN_A1}) alternatively from the local rate of expansion $\theta$ of the generating null-geodesics emanating from the origin. According to the Raychaudhuri equation in the linearized limit, 
\begin{eqnarray}
\frac{d\theta}{d\lambda} = -R_{ab}k^ak^b = - 8\pi \rho \left(u^a k_a\right)^2
\label{EQN_RS}
\end{eqnarray}
along the null-tangents $k^b=\frac{dx^b}{d\lambda}$, where $R_{ab} = T_{ab} - \frac{1}{2}g_{ab}T_c^c$ for a particle of mass with local mass-density $\rho$ and using a velocity four-vector $u^b=(\alpha^{-1},0,0,0)$ of local observers in the unperturbed Schwarzschild line-element of a mass $M$ centered at the origin. To fix an affine parameter $\lambda$ of the null-generators, we consider $u^ak_a=-1$, whereby $k^r=\alpha^2k^0= \alpha$, so that $d\lambda=\alpha^{-1}dr$, i.e., $\lambda$ represents the invariant length in the radial direction.

The area $A_f$ of the wave-front generated by the null-geodesics up to radius $R$ is perturbed by $\rho$ about $r_0<R$. According to (\ref{EQN_RS}), the wave front has a surface area 
\begin{eqnarray}
A_f(R,r_0) = \int_0^R \int \theta dA dr = \int_0^R\int_0^r \left[\int \frac{d\theta}{d\lambda}dA\right] d\lambda dr  = - 8\pi \int_0^R\int_0^r \left[ \int \rho  dA \right ] d\lambda dr.
\end{eqnarray}
Here, the minus sign is due to convergence of light rays signaling the presence of $m\delta(\lambda-\lambda_0)=\int \rho dA$ at $\lambda_0$. In the approximation $\lambda \simeq r$ for large separations, it follows that
\begin{eqnarray}
A_f(R,r_0) = 4\pi R^2 - 8\pi m \int_0^R \int_0^r \delta(\lambda-\lambda_0) d\lambda dr = 4\pi R^2-8\pi m (R-r_0),
\label{EQN_AR}
\end{eqnarray}
where $4\pi R^2$ refers to the unperturbed area arising from the integration constant in (\ref{EQN_RS}). Following (\ref{EQN_A1}), we may alternatively use a gauge that keeps the surface area of the wave-front fixed, whereby (\ref{EQN_AR}) gives rise to
a change in the projected surface area
\begin{eqnarray}
\Delta A_0 = -2\pi m \Delta r_0
\label{EQN_AR2}
\end{eqnarray}
due to a displacement $\Delta r_0$ of the world-line of $m$.

\begin{figure}[h]
\center{\includegraphics[scale=0.225]{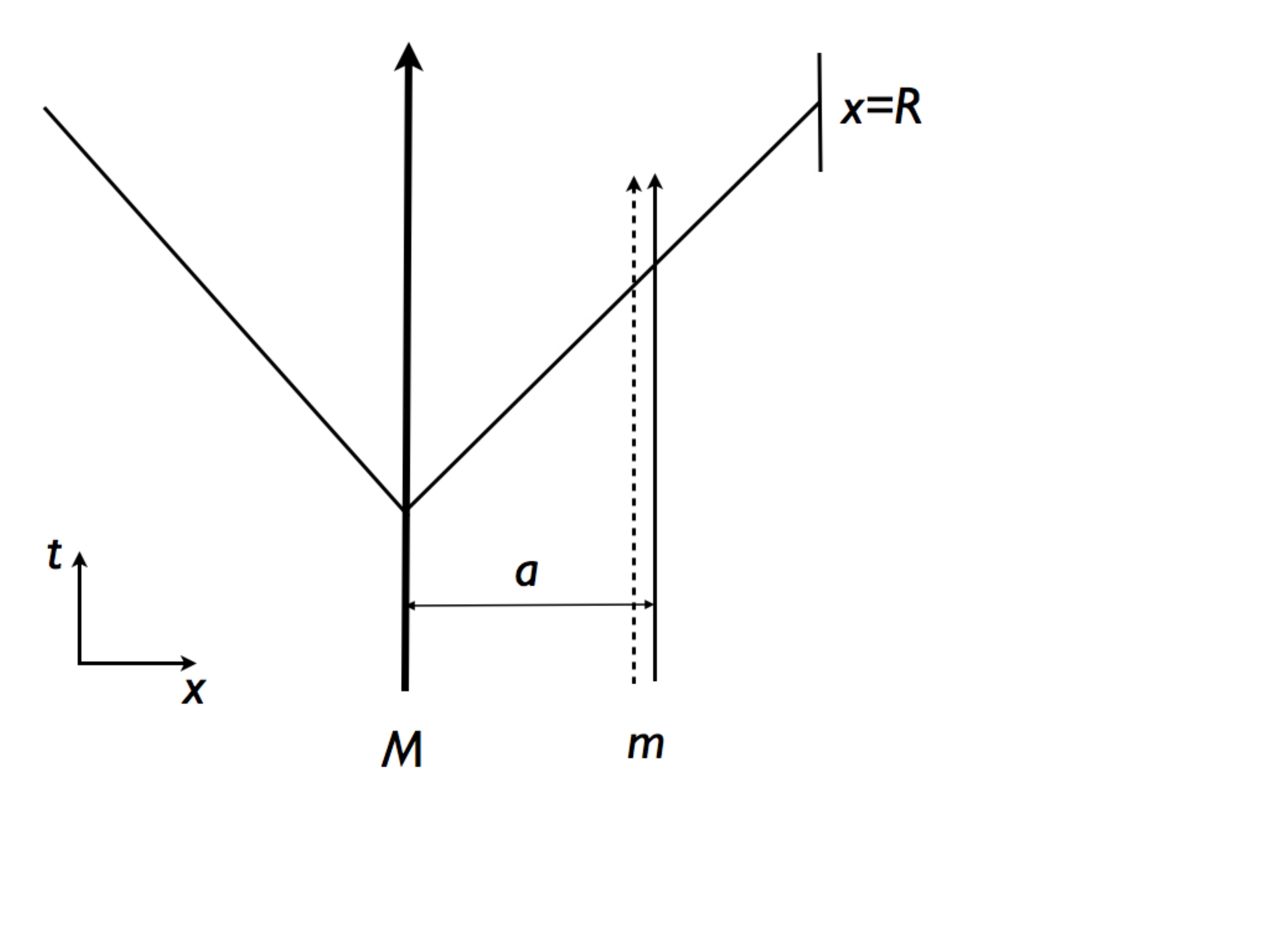}\includegraphics[scale=0.225]{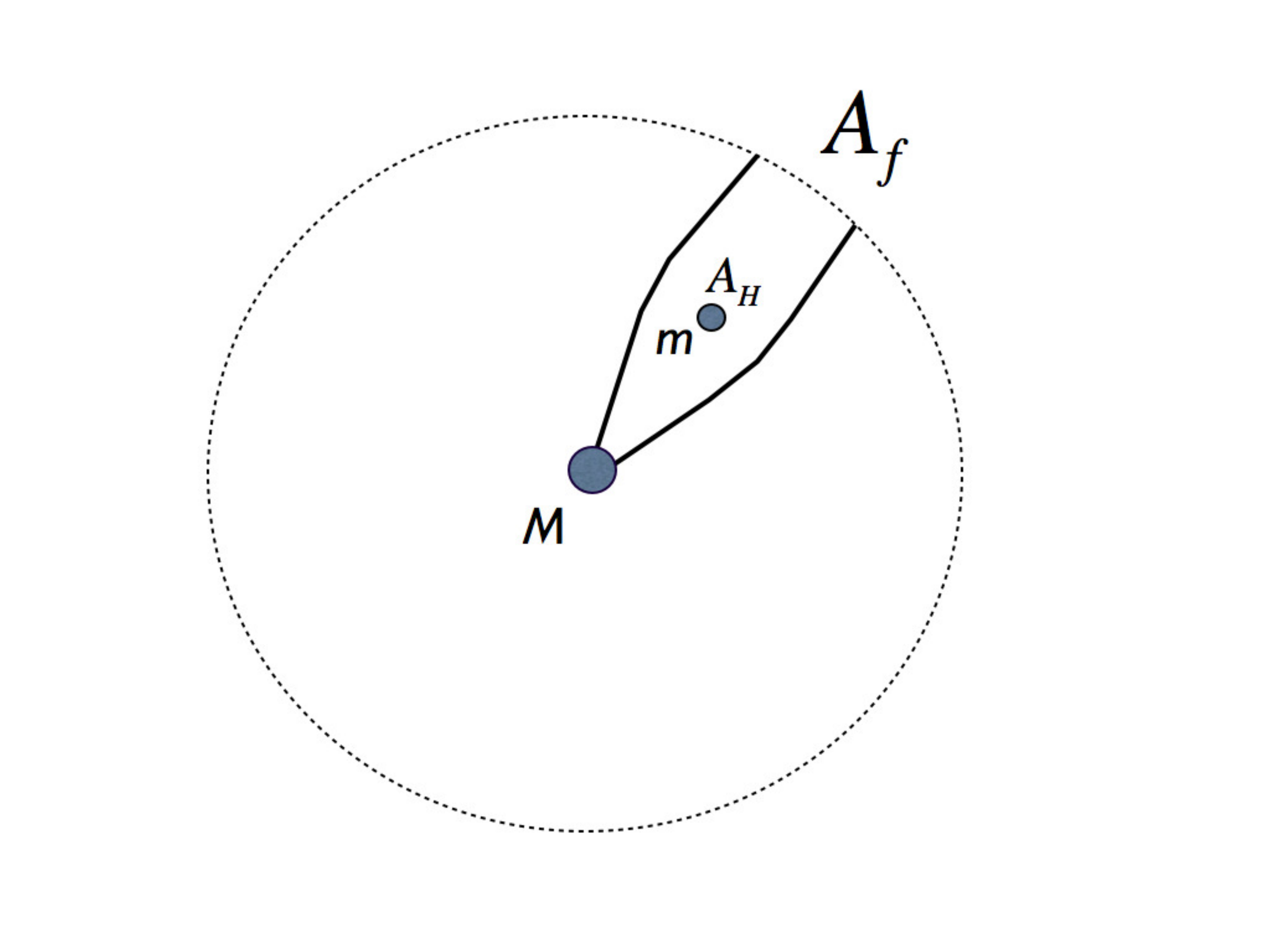}}
\caption{Shown is the derivation of Newton's force on $m$ at a distance $a$ from a central mass $M$ by application of Gibbs' procedure ($left$). Null rays generating a light cone emanating from the world-line of $M$ are subject to lensing by $m$ ($right$). The area $A_f$ of a wave front (a surface of constant phase) may be rendered finite by considering a cut-off radius $R>a$. As a null surface, the entropy of a light cone is the area of its projection onto the equatorial plane. This change in entropy thus imparted by lensing increases in strength following a virtual displacement of $m$ towards $M$. The result is an attractive entropic force by Gibbs' principle, independent of $R$, when the light cone is given an Unruh temperature proportional to $M/a^2$. If the particle $m$ is a black hole with an apparent horizon surface area $A_H$, changes $\delta A_H$ in its surface area are correlated to changes $\delta A_f$ in considering Gibbs' virtual displacement $\delta a$.}
\label{FIG_C10E}
\end{figure}
As illustrated schematically in Fig. \ref{FIG_C10E}, we next apply Gibbs' procedure, and consider the entropic force arising from a displacement of $m$ by $\Delta r_0$ in the presence of a mass $M$ at the origin. By (\ref{EQN_AR2}) and consistent with (\ref{EQN_A1}), the entropy change is considered to be the area change of the equatorial disk enclosed by a wave-front as $m$ crosses the light cone, 
\begin{eqnarray}
\Delta S = -2\pi m \Delta r_0.
\label{EQN_DS}
\end{eqnarray}
With the temperature $T=\frac{M}{2\pi r_0^2}$ set by the central mass in accord with (\ref{EQN_ST}), the resulting entropic force 
\begin{eqnarray}
F= T\frac{\Delta S}{\Delta r_0} = - \frac{M m}{R^2}
\label{EQN_F}
\end{eqnarray}
is Newton's law.

\section{Conclusions} 

Entropy is a central theme in gravity that is naturally attributed to null-surfaces as causal event horizons in relation to the mass-energy contained in spacetime. However, its manifestation in {\em forces}, notably Newton's law of gravitational attraction, is non-trivial when harmonizing it between black holes and particles in the Newtonian point mass limit.

We developed a systematic approach based on the exactly solvable problem of the entropic force between two black holes. From this example, we learn that {\em Newon's law is the adiabatic limit of gravitational interactions} at separations large to the Schwarzschild radii of the participating particles and two entropic constraints that any theory of entropic gravity force must satisfy: monotonicity of entropy and the Bekenstein entropy bound. 

We use these results to derive Newton's law for point particles as an entropic force by application of Gibbs' procedure applied to light cones, based on entropy variations induced by gravitational lensing. Here, an entropy is attributed to the area of the projection wave-front of constant phase onto the equatorial plane, and entropy changes by deformations due to lensing are similar to that seen, in pronounced form, in the Schwarzschild metric. It points to a complementary approach to the entropy of a black hole event horizon in terms of an integral quantity emerging from curvature in the surrounding spacetime; and its apparent relation to a projected surface suggests a potential relation to flux of unknown origin, as when calculating the net magnetic flux through a black hole event horizon (reviewed in \cite{van01}). 

In the Boyer-Lindquist embedding of a Schwarzschild black hole, flux lines crossing the event horizon pass from one sheet to the other. When flux is quantized, the surface area relevant for the total number of flux elements is the area of the surface of the event horizon projected onto the equatorial plane. The event horizon introduces maximal uncertainty between the ordering of flux lines in either sheet, and the corresponding entropy is given by the logarithm of all possible permutations between them. Viewed in one sheet, magnetic flux is illustrative: an electron falling along one flux line can annihilate with a positron along any other flux line upon reaching the event horizon, wherein the total number of field lines follows from the net magnetic flux through the event horizon.

Our geometric approach suggests that entropy is related to the causal distance between masses and wave fronts along light cones. This suggests that there may exist a complementary approach to the entropy of black hole event horizons, as an integral quantity emerging from the curvature of the exterior spacetime. Indeed, if the particle is a black hole of mass $m$, a virtual displacement following Gibbs' principle in Fig. 2 gives rise to a change in the surface area $\delta A_H$ of its apparent horizon in accord with (\ref{EQN_A2}) {\em and} a change $\delta A_f$ in the surface area of a screen defined by a wave front in accord with (\ref{EQN_AR}). Giving rise to the same Newtonian limit, we have
 \begin{eqnarray}
T_H\delta A_H + T\delta A_f = 0,
 \end{eqnarray}
where $T_H$ refers to the temperature on the apparent horizon of the black hole and $T$ is defined by (\ref{EQN_ST}) at the location of $m$. It expresses a concrete example of a property of a screen, here a wave front, as a function of the distribution of matter within.

In our approach, Planck's constant serves to relate surface area to entropy, as in the case of black hole event horizons, with area changes due to surface deformations attributed to lensing (\ref{EQN_RS}). Thus, (\ref{EQN_DS}) results from converting (\ref{EQN_AR2}) by the Planck length squared. In this process, Newton's constant in (\ref{EQN_AR2}) is cancelled. This provides some rationale for (\ref{EQN_DS}) and the appearance of Planck's constant therein as hypothesized by scaling in \citep{ver10} based alternatively on the Compton wave length of $m$. This observation leaves open further microphysical arguments, to explain the relation between lensing and Newton's law (\ref{EQN_F}). 

Our derivation of (\ref{EQN_AR2}) based on lensing and (\ref{EQN_ST}) differs from the scaling arguments in \citep{ver10}, in explicitly identifying the sign of the entropy change, i.e., increasing as the two particles get closer together. The above has in common with \citep{ver10} a focus on Newton's law between two massive point particles, in contrast to \citep{jac95}. While the Raychaudhuri equation is central to our derivation as in \citep{jac95}, our application to deriving Newton's law is entirely different with no need to invoke a Rindler observer.

Efforts to generalize entropic gravity by holography using time-like surfaces pose the challenge to calculate the distribution of entropy, that is generally variable across the screen, and its average surface density, that will be less than the extremal value for null-surfaces. Time-like surfaces involve fewer states than the total number of available states in a horizon surface, which scales exponentially with the mass-energy within times its linear size (e.g. \cite{cha11}). Detailed calculations of this kind are beyond the general considerations considered in this paper, e.g., to explore a possible relation of entropy to flux. It will require some microphysical model for holographic screens describing how matter is embedded in spacetime or how the two (and any accompanying interactions) emerge or merge (when matter falls into a black hole) in a low energy limit as sought after in \cite{ver10}.

{\bf Acknowledgments.} We gratefully thank constructive comments from stimulating discussions with Stephano Bolognesi, EE Chang-Young, Gerard 't Hooft, Kimyeong Lee, Piljin Yi, Erik Verlinde and constructive comments from the referee.

\end{document}